\documentclass[aps,prl,twocolumn,showpacs,floatfix,a4paper]{revtex4}
\usepackage{graphicx}
\usepackage[intlimits]{amsmath}

\begin{document}

\title{Duality between different geometries
of a resonant level in a Luttinger liquid}
\author{Moshe Goldstein}
\author{Richard Berkovits}
\affiliation{The Minerva Center, Department of Physics, Bar-Ilan
University, Ramat-Gan 52900, Israel}

\begin{abstract}
We prove an exact duality between the side-coupled and
embedded geometries of a single level quantum dot
attached to a quantum wire in a Luttinger liquid phase
by a tunneling term and interactions.
This is valid even in the presence of a finite bias voltage.
Under this relation the Luttinger liquid parameter $g$ goes
into its inverse, and transmittance maps onto reflectance.
We then demonstrate how this duality is revealed by the
transport properties of the side-coupled case.
Conductance is found to exhibit an antiresonance
as a function of the level energy, whose width vanishes
(enhancing transport)
as a power law for low temperature and bias voltage
whenever $g>1$,
and diverges (suppressing transport) for $g<1$.
On resonance transmission is always destroyed, unless
$g$ is large enough.
\end{abstract}


\maketitle

\emph{Introduction.}---
Understanding the behavior of strongly correlated systems
has been one of the central themes of condensed
matter physics in recent years.
Of these, one-dimensional systems stand out as a clear example
of non Fermi liquid behavior.
When no symmetry is spontaneously broken, the low energy
physics of those systems is governed by the bosonic
Luttinger liquid (LL) theory
\cite{bosonization}.
This description applies to various experimental realizations,
including semiconducting quantum wires, metallic nanowires,
and carbon nanotubes;
it is also related to the physics of the edges of
quantum Hall systems \cite{chang03}
and spin quantum Hall systems \cite{koenig08}.
An important question, from both fundamental and applicative
perspective, is the effect of (randomly or intentionally introduced)
impurities on such systems.
Particularly interesting are \emph{dynamic} impurities, e.g.,
resonant levels which can fluctuate between the occupied
and unoccupied states.
They can be realized, among other possibilities,
as semiconducting quantum dots, metallic grains,
or carbon nanotubes or buckyballs.
Indeed, much effort has been attracted to the understanding
of their effects on transport
\cite{bosonization,kane92,furusaki93,fendley95,weiss95,
furusaki98,auslaender00,postma01,nazarov03,polyakov03,komnik03,lerner08}
as well as thermodynamic
\cite{furusaki02,lehur05,sade05,wachter07,weiss08,bishara08,goldstein08}
properties.

\begin{figure}[b]
\includegraphics[width=6cm,height=!]{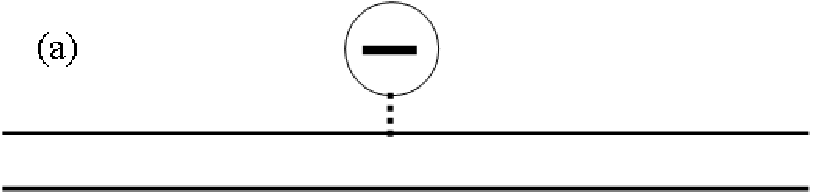}
\vskip 0.2cm
\includegraphics[width=6cm,height=!]{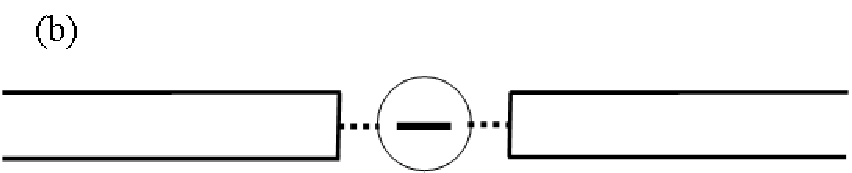}
\caption{\label{fig:geometry}
The geometry of the system:
(a) the original side-coupled configuration;
(b) the dual embedded configuration.
}
\end{figure}

An important insight into strongly interacting theories
is provided by dualities, i.e., mappings between the properties
of a system and those of a different system, usually with
reversed coupling strengths. In condensed matter physics,
this goes back to the famous Kramers-Wannier duality of the Ising model. 
Another example, in the context of this work, is the duality
between the strong- and weak-coupling limits of a static (potential)
impurity in a LL \cite{kane92}.
In this work we find a different kind of
duality for a level coupled to a LL: an equivalence
between the side-coupled and embedded geometries depicted
in Fig.~\ref{fig:geometry},
valid even in the presence of a finite bias voltage.
In this mapping,
the LL parameter $g$ goes onto $1/g$ and transmittance goes
into reflectance, but the strength of the
level-lead coupling is \emph{unchanged}: a strongly (tunnel-) coupled
level is mapped onto a strongly coupled level, and vice-versa.
In the following, after proving this result, we
demonstrate its power by characterizing
the transport properties of the side-coupled system for arbitrary
strength of the electron-electron interaction,
which, to the best of our knowledge,
have been previously discussed only for weak interaction
\cite{lerner08}.

\emph{Model.}---
The system depicted in Fig.~\ref{fig:geometry}(a)
is described by the Hamiltonian
$H=H_W+H_D+H_T$,
where, for spinless particles (spin effects will be discussed later):
\begin{equation}
 H_W = \frac{v}{2\pi}
	\int_{-\infty}^{\infty} \left\{ \frac{1}{g} [\partial_x \Theta(x)]^2 +
	g [\partial_x \Phi(x)]^2  \right\} \text{d}x ,
 \label{eqn:hw}
\end{equation}
is the lead bosonized Hamiltonian, expressed in terms of the bosonic
fields $\Theta(x)$ and $\Phi(x)$, obeying the commutation relation
$[\Theta(x),\Phi(y)]=-\text{i}\frac{\pi}{2}\text{sgn}(x-y)$, and where $g$ is
the interaction parameter
($g<1$ for repulsion, $g>1$ for attraction)
and $v$ is the velocity of excitations
\cite{bosonization}.
The level Hamiltonian is $H_D=\varepsilon_0 d^{\dagger} d$, with
$d$ the level Fermi operator, and $\varepsilon_0$ its energy.
The level and the lead are connected by a tunneling term
(effects of level-lead interaction will be considered momentarily):
\begin{equation}
 H_T = t_0 d^{\dagger} \left[ \psi_{+}(0) + \psi_{-}(0) \right] + \text{H.c.}
 \label{eqn:ht}
\end{equation}
Here $t_0$ is the tunneling matrix element, and
the lead right (left) moving Fermi operators can be expressed
in terms of the bosonic fields through
$\psi_{\pm}(x)=
\sqrt{\frac{D_0}{2\pi v}} \chi_{\pm} \text{e}^{\text{i}\Phi_{\pm}(x)}$,
where $\Phi_{\pm}(x) = \pm \Theta(x) - \Phi(x)$
are chiral right (left) moving fields
obeying $[\Phi_\alpha(x),\Phi_\beta(y)] =
\alpha\text{i}\pi\delta_{\alpha\beta}\text{sgn}(x-y)$
and $\chi_{\pm}$ are Majorana Fermions
($D_0$ is the bandwidth).

\emph{Duality.}---
We now turn to the derivation of our central result: the duality
between the side-coupled and embedded geometries.
Let us define two new bosonic fields,
$\theta(x) \equiv \Theta(x)/\sqrt{g}$
and $\phi(x) \equiv \sqrt{g}\Phi(x)$,
in terms of which the interaction parameter $g$
is eliminated from the $H_W$
but introduced into $H_T$.
We can also use these fields to write down \emph{decoupled}
(Bogolubov-transformed) right and left
moving fields $\phi_{\pm}(x) = \pm \theta(x) - \phi(x)$.
We then apply a unitary transformation,
$\tilde{H} = \mathcal{U}^{\dagger} H \mathcal{U}$, where
$\mathcal{U}=\text{e}^{\text{i} (g-1) \left( d^{\dagger}d - \frac{1}{2} \right)
\phi(0) / \sqrt{g} }$.
The transformed Hamiltonian is similar to the original one, except for
the addition of a term of the form
$(g-1) v \left( d^{\dagger}d - \frac{1}{2} \right) \partial_x \theta(0) / \sqrt{g}$,
as well as a modification of the Fermi operators at the origin to
$\tilde{\psi}_{\pm}(0) = \sqrt{\frac{D_0}{2\pi v}} \chi_{\pm}
\text{e}^{\text{i} \sqrt{g} (\pm \theta(0) - \phi(0)) }$.
The Hamiltonian now takes the form
$\tilde{H}=\tilde{H}_W+\tilde{H}_D+\tilde{H}_T$, with $\tilde{H}_D=H_D$, and:
\begin{align}
 \tilde{H}_W =& \sum_{\alpha=\pm} \frac{v}{4\pi}
	\int_{-\infty}^{\infty} [\partial_x \phi_\alpha(x)]^2 \text{d}x,
 \label{eqn:hw_tilde}
 \\
 \tilde{H}_T =&
	t_0 d^{\dagger}
	\sum_{\alpha=\pm}
	\sqrt{\frac{D_0}{2\pi v}} \chi_\alpha
	\text{e}^{\text{i} \sqrt{g} \phi_\alpha(0) }
	+ \text{H.c.} \nonumber \\ &
	+ (g-1) \pi v \left( d^{\dagger}d - \textstyle\frac{1}{2} \right)
	\sum_{\alpha=\pm} \frac{\alpha \partial_x \phi_\alpha(x)}{2 \pi \sqrt{g}},
 \label{eqn:ht_tilde}
\end{align}
This is the Hamiltonian of \emph{two} \emph{chiral}
LLs (corresponding to the decoupled left and right movers in the
original model) with LL parameter $\tilde{g} = 1/g$, which are symmetrically
coupled to a level by both
a tunneling term of the same amplitude $t_0$, and a local charging interaction
of strength $\tilde{U}_F = (g-1) \pi v$.
But coupling a level to a chiral LL is known to be
equivalent to coupling it to the edge of a non-chiral LL
\cite{fabrizio95}.
This is achieved here by
defining $\tilde{\phi}_{\pm}(x) = \phi_{+}(\pm x)$
[$\tilde{\phi}_{\pm}(x) = \phi_{-}(\mp x)$]
for $x>0$ [$x<0$] as the decoupled right and left moving
fields in the new right [left] non-chiral lead.
Thus, this result proves the celebrated duality symmetry:
a level side-coupled to a LL [Fig.~\ref{fig:geometry}(a)]
is equivalent a level embedded (in a left-right symmetric manner)
between the edges to two LLs
[Fig.~\ref{fig:geometry}(b)].
The level energy $\varepsilon_0$ as well as the tunneling matrix element $t_0$
remain unchanged, but $g$ is transformed to $1/g$.
In addition, a local level-wire interaction must be included
\cite{fn:uf,fn:cutoff}.

How do the measurable properties of the system map under the duality?
It is easy to see that the level population
and its correlation functions (determining the dynamic capacitance), as
well as other thermodynamic properties of the level (e.g., its contribution
to the entropy and specific heat)
remain invariant under all the transformations performed,
and are thus equal for the side-coupled and embedded geometries.
Transport properties, however, do change.
To see this, let us use a Landauer-type formalism
\cite{egger98,safi99,fn:gx}.
Attaching the side-coupled system
at $x=\pm L/2$ to reservoirs at potentials $\pm V_0/2$
is equivalent to imposing the boundary conditions
$\cosh(\varphi) \rho_\pm \left( \mp L/2 \right) +
\sinh(\varphi) \rho_\mp \left( \mp L/2 \right) =
\pm \frac{eV_0}{4 \pi v}$
on the average 
decoupled right and left moving densities
$\rho_\pm(x) = \langle \pm \partial_x \phi_\pm(x) \rangle / (2 \pi)$,
with $\exp(- 2 \varphi) \equiv g$.
Summing these two equations, and using current conservation,
$\rho_+(-L/2)-\rho_-(-L/2)=\rho_+(+L/2)-\rho_-(+L/2)$, we get
$\rho_+(+L/2) = -\rho_-(-L/2)$.
Substituting this relation back in the
boundary conditions, and rewriting them in terms of the dual variables
(in terms of which left movers at $x>0$ become right movers at $x<0$
and vice versa,
while $\tilde{\varphi}=-\varphi$),
we find that the same expression holds for the embedded configuration:
$\cosh(\tilde{\varphi}) \tilde{\rho}_\pm \left( \mp L/2 \right) +
\sinh(\tilde{\varphi}) \tilde{\rho}_\mp \left( \mp L/2 \right) =
\pm \frac{eV_0}{4 \pi v}$.
In addition, since the average current can be written as
$I = \frac{e v \sqrt{g}}{2}
\left[\rho_+(0^+) + \rho_+(0^-) - \rho_-(0^+) - \rho_-(0^-) \right]$,
while the average voltage drop at $x=0$ is given by
$V = \frac{\pi v}{e \sqrt{g}}
\left[\rho_+(0^+) + \rho_-(0^+) - \rho_+(0^-) - \rho_-(0^-) \right]$,
we see that $I$ and $G_0 V$ are interchanged under the duality
transformation ($G_0 =e^2/h$ is the quantum conductance).
Now, in a steady state $\rho_\pm(x)$ are separately
constant for $x>0$ and $x<0$.
Thus, subtracting the boundary condition equations we have
$I+G_0 V = G_0 V_0$, which leads to the relation
$\tilde{I}+I=G_0 V_0$
between the currents
in the two geometries.
This result is physically clear:
when the level-lead coupling is weak, conductance is good for the
side-coupled system (no scattering), but is bad for the embedded one (no
tunneling) and vice versa.

It should be noted that one could have also included a local interaction
between the electrons in the level and those in the lead in our original system.
This would amount to adding to the Hamiltonian $H$ the term
\begin{equation}
 H_{U} = \left( d^{\dagger}d - \textstyle\frac{1}{2} \right)
	\left\{ \frac{U_F} {\pi} \partial_x \Theta(0)
	+ U_B  \left[ \psi_{L}^{\dagger}(0) \psi_{R}(0) + \text{H.c.} \right]
	\right\},
 \label{eqn:hu}
\end{equation}
with $U_{F(B)}$ the strength of the forward (backward) interaction.
Repeating all the transformations, we again obtain the same dual description in
terms of the embedded level, with two modifications:
(i) the strength of the local level-wire interaction in the embedded geometry
is now $\tilde{U}_F = \pi v (g-1) + g U_F$  \cite{fn:uf};
(ii) there will be an additional term, of the form
$U_B  \left( d^{\dagger}d - \frac{1}{2} \right) \left[ \tilde{\psi}_{L}^{\dagger}(0) \tilde{\psi}_{R}(0) + \text{H.c.} \right]$.
Although such a term is not usually included in the bare Hamiltonian of
the embedded geometry, it is nevertheless generated by virtual processes
in which an electron from one lead hops into the level and then into the other
lead and vice-versa.
Thus, this term does not add any new
physics into the system, and only changes the results presented below
quantitatively and not qualitatively (changing, e.g., the exact shapes
of phase boundaries but not their weak- and strong-coupling limits,
and affecting prefactors but not exponents in power-laws).
Finally, it may be added that the duality can be obtained
by comparing the Coulomb gas expansions for the two systems.
Details will be given elsewhere \cite{long_version,fn:gx}.

\begin{figure}
\includegraphics[width=9cm,height=!]{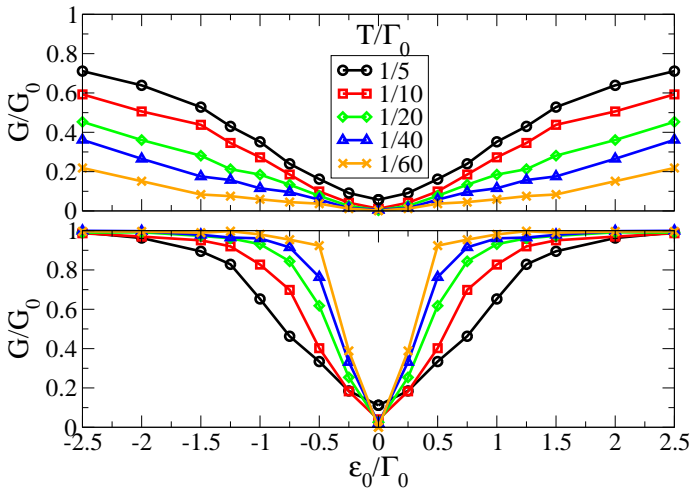}
\caption{\label{fig:resonance}
(Color online)
Linear conductance of the side-coupled geometry
as a function of the level energy at different temperatures
for $g=2/3<1$ (upper panel) and $g=3/2>1$ (lower panel).
See the text for further details.
}
\end{figure}

\emph{Transport properties.}---
In the rest of this paper we will study the transport
properties of the side-coupled system for arbitrary interaction strengths,
and show how the duality with the embedded case is revealed.
Let us start from a qualitative description.
The embedded geometry should behave similarly to the case
of a LL with two barriers tuned to resonance
\cite{kane92,furusaki93,nazarov03,polyakov03,komnik03}. Then, for not
too strong interactions the conductance is predicted to have a
resonance lineshape as a function of $\varepsilon_0$.
Without interactions ($g=1$) the lineshape is Lorentzian,
and its width saturates at low temperature $T$
to $\Gamma_0=\pi |t_0|^2 \nu_0$, $\nu_0=1/(\pi v)$
being the local density of states.
For $g<1$ the width decreases as the temperature is lowered,
suppressing conductance for $\varepsilon_0 \ne 0$, while
for $g>1$ the width increases, so that transport becomes
perfect at low enough $T$.
By the transmission-reflection duality,
we expect to have an anti-resonance for the
side-coupled geometry. From the above,
for $g=1$ the lineshape is Lorentzian, and its width
saturates to $\Gamma_0$ at low temperature,
as one can immediately verify by a direct calculation.
However, by the $g\leftrightarrow1/g$ correspondence,
here for $g<1$ the width should \emph{increase} as $T$ 
is lowered, whereas for $g>1$ it should \emph{decrease}.
These expectations are borne out by Monte-Carlo calculations on
a Coulomb-gas representation, to be discussed
elsewhere \cite{long_version}. However, as an illustration we plot
some of the results in Fig.~\ref{fig:resonance}.
Thus, in both geometries conductance is suppressed for $g<1$,
and enhanced for $g>1$,
but this is realized by \emph{opposite} lineshapes in the two configurations.
As we now show, these considerations are supported by
direct analysis of the side-coupled problem.

We will first consider the limit of weak  level-lead coupling
for arbitrary values of $g$.
Then, let us apply the transformation
$H^\prime = \mathcal{V}^\dagger H \mathcal{V}$, where
$\mathcal{V} = \text{e}^
{- \text{i} \delta_F \left( d^{\dagger}d - \frac{1}{2} \right) \phi(0)/\sqrt{g} }$,
$\delta_F \equiv g U_F/(\pi v)$.
This eliminates the forward interaction term from the Hamiltonian,
at the cost of modifying the tunneling term.
In terms of the dimensionless parameters
$y_t \equiv \sqrt{\Gamma_0 \xi/(2 \pi)}$ and $y_B \equiv U_B /(4 \pi v)$
($\xi = 1/D$ is a short time cutoff),
the level-lead coupling terms in $H^\prime$ now read:
\begin{align}
 \label{eqn:ht_prime}
 H^\prime_T = & \frac{y_t}{\xi} d^{\dagger} \sum_{\alpha=\pm} \chi_\alpha
	\text{e}^{\text{i} \sum_{\alpha^\prime = \pm}
	K_{\alpha,\alpha^\prime} \phi_{\alpha^\prime}(0)}
	+ \text{H.c.}, \\
 \label{eqn:hu_prime}
 H^\prime_U = & \frac{2 y_B}{\xi} \left( d^{\dagger}d - \textstyle\frac{1}{2} \right)
	\chi_+ \chi_- \text{e}^{\text{i} \sqrt{g} [\phi_+(0)-\phi_-(0)] }
	+ \text{H.c.},
\end{align}
where
$K_{\alpha,\alpha^\prime} \equiv
(1-\delta_F)/(2\sqrt{g}) + \alpha \alpha^\prime \sqrt{g}/2$.
The scaling dimensions of $H^\prime_T$ and $H^\prime_U$ are thus
$(K_{1,1}^2+K_{1,-1}^2)/2$ and $g$, respectively.
In addition, the vertex operators
$V^\alpha_a =
\mbox{$:\exp \left[ \text{i} a \phi_\alpha \right]:$}$
obey the operator product expansion \cite{bosonization}
$V^+_a (z^\prime) V^+_b (z) \sim (z^\prime-z)^{a b} V^+_{a+b}(z)$
($a \neq -b$),
$V^+_a (z^\prime) V^+_{-a} (z)
\sim (z^\prime-z)^{1 - a^2} \text{i} a \partial_z \phi_+(z)$,
and similarly for $V^-_a$.
Substituting this in Cardy's general formulas
gives the RG equations to second order in
$y_t$ and $y_B$ \cite{cardy}:
\begin{align}
 \label{eqn:rg_yt}
 \frac {\text{d}y_t} { \text{d} \ln \xi} & =
 \left[ 1 - \frac{g}{4} - \frac{(1 - \delta_F)^2}{4g} \right] y_t
   + y_t y_B \\
 \label{eqn:rg_df}
 \frac {\text{d}\delta_F} { \text{d} \ln \xi} & =
 ( 1 - \delta_F ) (4 y_t^2 +y_B^2)\\
 \label{eqn:rg_yb}
 \frac {\text{d}y_B} { \text{d} \ln \xi} & =
 (1 - g) y_B + y_t^2
\end{align}
Off-resonance ($\varepsilon_0 \ne 0$), the flow of $y_t$ is
stopped as soon as $\xi \sim \varepsilon_0$. From this point on,
the level is locked into one of its two possible states
(occupied or empty, depending the sign of $\varepsilon_0$),
and the only RG equation left is
$\text{d}y_B/\text{d} \ln \xi = (1 - g) y_B$.
This simply means that off-resonance the level acts as a potential
scatterer, whose strength $\sim t_0^2/\varepsilon_0$ for large
enough $\varepsilon_0$.
From these equations we see that $U_B$ (which is generated by terms
of second order in $t_0$ even if not present in the original Hamiltonian)
is in general relevant for $g<1$ and irrelevant for $g>1$, as
expected for a backscattering term \cite{kane92}.
$t_0$ (which directly affects the low energy physics only for $\varepsilon_0$=0)
is relevant for $g_{-}<g<g_{+}$, where $g_\pm$ are
the solutions of $g + [ 1 - (g U_F)/(\pi v) ]^2/g = 4$.
Whenever any of these two terms is relevant, scattering induced
by the level destroys conductance for small $T$ and $V_0$.

For strong level-lead coupling,
the forward scattering rapidly converges to its fixed point value
$U_F = \pi v / g$ [cf.\ Eq.~(\ref{eqn:rg_df})].
On resonance, the hopping term in the Hamiltonian is
more relevant than the backscattering term,
so we can concentrate on it in the strong-coupling limit.
After the transformation $\mathcal{V}$ described above it becomes
$H^\prime_T = 4 y_t \xi^{-1} S_x
\cos \{\sqrt{g}[\phi_+(0)-\phi_-(0)]/2\}$,
where we have defined the spin variables $S_{+} \equiv d^\dagger$,
$S_{-} \equiv d$, and $S_z \equiv d^\dagger d - 1/2$
\cite{fn:klein_factors}.
Since $S_x$ commutes with $H^\prime_T$,
it assumes one of its possible values ($\pm 1/2$).
Then $H^\prime_T$ take the form of a potential backscattering
term, but with $g$ replaced by $g/4$.
From the known behavior of the latter problem \cite{kane92} we can infer
that the strong $y_t$ (suppressed transmission) limit
is stable for $g<4$ and unstable for $g>4$.
Off resonance only the $H^\prime_U$ term is important at low energies
(below $\varepsilon_0$, where $H^\prime_T$ is frozen). 
In the spin notation it becomes
$H^\prime_U = 2 y_B \xi^{-1} S_z \cos \{\sqrt{g}[\phi_+(0)-\phi_-(0)]/2\}$,
where $S_z = 1/2$ for $\varepsilon_0<0$ and vice versa.
Now $H^\prime_U$ behaves like a usual potential backscattering
so the strong $y_B$ limit is stable for $g<1$, unstable for $g>1$.

\begin{figure}
\includegraphics[width=9cm,height=!]{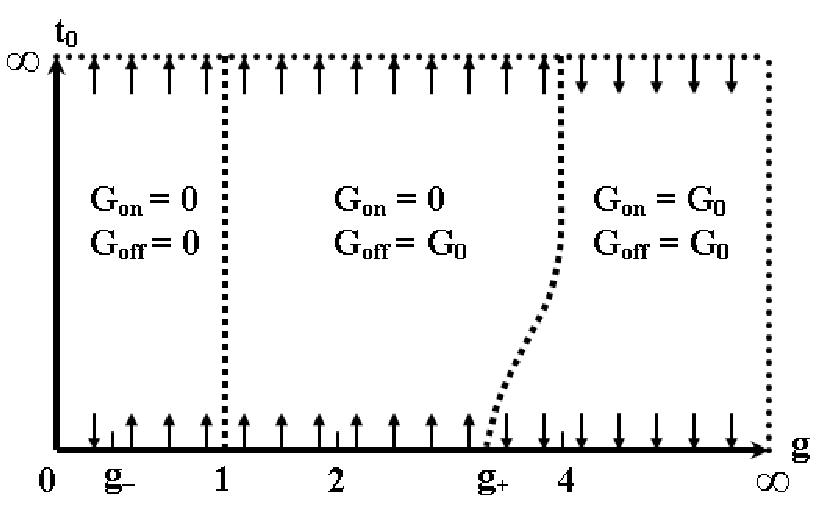}
\caption{\label{fig:phase_diagram}
Zero temperature conductance phase diagram and on-resonance RG flow,
projected on the $g$-$t_0$ plane.
See the text for further details.
}
\end{figure}

Taking all these results together, we can find the phase diagram
of the system, plotted in Fig.~\ref{fig:phase_diagram}.
There are three phases at $T=0$:
(i) for $g<1$ conductance is suppressed both on and off resonance,
by a widening anti-resonance.
(ii) for $g>1$ but not too large,
we obtain a narrowing anti-resonance, so that at low temperatures
transport is perfect everywhere except for $\varepsilon_0=0$.
(iii) for large enough $g$ [$g>4$ for large $t_0$,
$g>g_{+}+C_{+}\sqrt{\Gamma_0/(2 \pi D_0)}+O(\Gamma_0/D_0)$ with
$C_{+}=4g_{+}\sqrt{2(4-g_{+})}/(g_{+}-g_{-})$ for small $t_0$]
perfect conductance is attained even on-resonance.
Moreover, concentrating on phases (i) and (ii)
(i.e., not too strong attraction),
the anti-resonance width scales as $\Lambda^{1-1/g}$,
where $\Lambda=\max(T,V_0,\pi v/L)$ is the largest infrared cutoff,
and where the power is determined by the scaling dimension of
the leading correction to the large $y_B$ limit,
i.e., tunneling across a barrier at $x=0$.
In the vicinity of $\varepsilon_0=0$
the conductance behaves as $\Lambda^{2(1/g-1)}$,
while for large $|\varepsilon_0|$ it deviates from the perfect value of $e^2/h$
by a correction proportional to $\Lambda^{2(g-1)}$
(determined by the scaling dimension of $H^\prime_U$) 
\cite{fn:g_lt_gm}.
As they should, all these results obey the duality
relation with those for resonant tunneling \cite{kane92}.

There are two cases in which the side-coupled system can be analyzed by
different methods (valid for arbitrary level-lead coupling),
and compared with similar calculations for the embedded geometry:
(i) The limit of weak electron-electron interactions ($g$ near 1),
which has been recently addressed by fermionic perturbative
(in the electron-electron interaction) RG methods \cite{lerner08},
previously employed to study of the
embedded configuration \cite{nazarov03,polyakov03};
(ii) Exact solution by refermionization at $g=2$,
in analogy with the embedded case at $g=1/2$ \cite{komnik03}.
These results can be shown to confirm both the general analysis given
above as well as the duality relation \cite{long_version}.

\emph{Including spin.}---
Finally we note that the derivation of the duality symmetry can be
easily extended to the spinful case, i.e., the Anderson impurity
model coupled to a LL, relevant for the problem of the Kondo effect
in a LL \cite{furusaki05}.
Both the charge and spin LL parameters \cite{bosonization} transform as
$g_\lambda \leftrightarrow 1/g_\lambda$
[$\lambda=c$ ($s$) for charge (spin)].
The strength of the charge and spin
level-lead interaction in the embedded case is
$\tilde{U}_{F,\lambda}=\pi v_\lambda (g_\lambda-1)/2 + g_\lambda U_{F,\lambda}$.
This means that for $g_s \ne 1$, implying spin anisotropy in the
wire, we will need to include spin-anisotropic level-lead
interaction, marked by nonzero $\tilde{U}_{F,s}$. 
Similar extension to a many level dot is also possible.

\emph{Conclusions.}---
To conclude, we have shown that for a level coupled to a LL lead there
exists a duality symmetry between the side-coupled and embedded geometries,
and examined it through a study of transport properties
in the two systems. As we have seen, the conductance lineshape
behaves in the opposite way in the two configurations, only to lead to the
same final result: at low temperature, transport is suppressed for
$g<1$ by a narrowing resonance (widening anti-resonance)
for the embedded (side-coupled) configuration, and vice-versa for $g>1$.
These findings have an important implication on experiments
\cite{auslaender00,postma01}:
since in reality electrons repel each other, physical realizations
of the systems discussed are limited to $g<1$.
However, the physics of attractive interactions ($g>1$) in each
geometry now becomes experimentally accessible
through investigation of the behavior for $g<1$ in the dual system.

\begin{acknowledgments}
We would like to thank Y. Gefen, I.V. Lerner, A. Schiller, and I.V. Yurkevich
for useful discussions.
M.G. is supported by the Adams Foundation Program of the Israel Academy
of Sciences and Humanities.
Financial support from the Israel Science Foundation (Grant 569/07) is
gratefully acknowledged.
\end{acknowledgments}


\begin{thebibliography}{99}
\bibitem{fn:uf}
Actually, Eqs.~(\ref{eqn:rg_yt})-(\ref{eqn:rg_df}) imply that
for not too strong interactions $y_t$ is relevant.
Then $U_F$ flows to its fixed point value,
so its precise bare value does not affect qualitatively
the low energy physics.
\bibitem{fn:klein_factors}
The Majorana fermions can be absorbed into these definitions to make
the spin variables commute with the lead degrees of freedom \cite{komnik03}.
\bibitem{fn:g_lt_gm}
For $g<g_{-}$ and small $t_0$, $U_B$ might become dominant even
on resonance, possibly modifying this behavior.
\end{thebibliography}
\end{document}